\documentclass[adp,fleqn]{w-art}
\usepackage{times}
\usepackage{w-thm}
\usepackage[]{graphicx}
\usepackage{longtable,ltxtable} 
\usepackage{epstopdf}
\chardef\bslash=`\\ 
\newcolumntype{L}{>{$}l<{$}}

\newcommand{\bs}[1]{\boldsymbol{#1}}
\newcommand{\comm}[2]{\left[#1,#2\right]}

\newcommand{\ket}[1]{\left|#1\right\rangle}

\newcommand{\up}{\uparrow}
\newcommand{\dw}{\downarrow}
\def\ie{{ i.e.},\ }
\def\eg{{ e.g.}\ }

\begin{document}
\DOIsuffix{theDOIsuffix}
\Volume{12}
\Issue{1}
\Copyrightissue{01}
\Month{01}
\Year{2008}
\pagespan{1}{}
\Receiveddate{16 June 2008}
\Reviseddate{8 August 2008}
\Accepteddate{20 August 2008 by U. Eckern}



\title[DMRG studies of critical SU($N$) spin chains]{DMRG studies of critical SU($\bs{N}$) spin chains}

\address[\inst{1}]{Institut f\"ur Theorie der Kondensierten Materie,
  Universit\"at Karlsruhe, 76128 Karlsruhe, Germany}
\address[\inst{2}]{Institut f\"ur Nanotechnologie, Forschungszentrum
   Karlsruhe, 76021 Karlsruhe, Germany}
\author[M. F\"uhringer et al.]{Max F\"uhringer$^{1}$}
\author[]{Stephan Rachel%
$^{1}$}
\author[]{Ronny Thomale$^{1}$}
\author[]{Martin Greiter$^{1}$}
\author[]{Peter Schmitteckert$^{2}$}

\begin{abstract}
  The DMRG method is applied to integrable models of antiferromagnetic
  spin chains for fundamental and higher representations of SU(2),
  SU(3), and SU(4).  From the low energy spectrum and the entanglement
  entropy, we compute the central charge and the primary field scaling
  dimensions. These parameters allow us to identify uniquely the
  Wess--Zumino--Witten models capturing the low energy sectors of the
  models we consider.
\end{abstract}
\maketitle                   





With the rise of quantum mechanics in the late 20's of the last
century~\cite{schroedinger26ap109,Heisenberg27zp172}, quantum
magnetism emerged as a predominant area of research in theoretical
condensed matter physics.  This was to a significant part induced by
the notion of the electron spin, \ie the magnetically sensitive,
internal degree of freedom of electrons, in the early 20's, which
rendered the classical picture insufficient.  In contrast to 
orbital angular momentum, which is
quantized in integer units of $\hbar$ in accordance with the spatial
rotation group SO(3), the internal spin is in accordance with the Lie
group SU(2) quantized in integer units of $\hbar$ (the generators of
both groups are identical as SU(2) is locally isomorphic to SO(3)).
Ever since the invention of the Bethe ansatz in 1931 as a method to
solve the $S=1/2$ Heisenberg chain with nearest-neighbor
interactions~\cite{bethe31zp205}, spin models in (1+1) dimension, \ie
quantum spin chains, have been a most rewarding subject of study.
Bethe's work eventually led to the discovery of the Yang-Baxter
equation in 1967~\cite{yang67prl1312} and provides the foundation of
the field of integrable models.  The notion of integrability rendered
a plethora of models amenable to exact and often rather explicit
solution~\cite{faddeev82proc,KorepinBogoliubovIzergin93}. Quantum spin
chains possess rich and deeply complex physical properties.  For
example, it took several decades until Faddeev and
Takhtajan~\cite{faddeev-81pla375} discovered in 1981 that the
elementary excitations of the $S=1/2$ Heisenberg chain solved by Bethe
carry spin $1/2$ and not, as previously assumed, spin 1.  The
excitations of the spin $1/2$ chain hence provide an instance of
fractional quantization, as the Hilbert space for the chain is spanned
by spin flips, which carry spin 1.

Several new aspects, both phenomenological and technical, emerge when
the spins transform under higher representations of SU(2).  In
particular, Haldane proposed in 1983 that half-integer spin chains are
generically gapless, whereas integer spin chains possess a gap in the
excitation
spectrum~\cite{haldane83pla464,haldane83prl1153,affleck90proc,Fradkin91}.
This leads to strikingly different behavior in the magnetic
susceptibility at low temperatures.  A gap leads to exponentially
decaying spin-spin correlations and as such to a vanishing
susceptibility at temperature $T=0$. In contrast, a gapless spectrum
is generically associated with correlations which decay as a power law
with the distance, and a
finite susceptibility at low temperatures.  Haldane's at that time
astonishing prediction was confirmed experimentally in
$S=1$ chains~\cite{buyers-86prl371,renard-87epl945,ma-92prl3571}.

Yet another generalization of quantum spin chains is to enlarge the
spin symmetry group from SU(2) to SU($N$)~\cite{Cornwell84vol2}.
Among those, the group SU(3) plays a special role, as both color and
flavor symmetries in particle physics provide instances.
As the electron spin transforms according to the fundamental (up/down)
doublet representation of SU(2), an internal ``color'' degree of
freedom of quarks transforms according to the three dimensional
fundamental representation of SU(3), and thus can be assigned a
quantum number taking the values blue, red, and
green.

\begin{vchfigure}[b]
  \centering
\setlength{\unitlength}{10pt}
  \begin{picture}(20,9)(20,-1)
\put(26,0){\line(0,1){6}}
\multiput(26,1)(0,2){3}{\circle*{0.5}}
\put(25,6){\makebox(1,1){$S^z$}}
\put(27.6,0.5){\makebox(2,1){$-1~~\ket{\dw\dw}$}}
\put(28.6,2.5){\makebox(5,1){$\phantom{+}0~~\tfrac{1}{\sqrt{2}}\bigl(
\ket{\up\dw}+\ket{\dw\up}\bigr)$}}
\put(27.6,4.5){\makebox(2,1){$+1~~\ket{\up\up}$}}
\put(26.6,-2.5){\makebox(2,1){a) Spin 1 representation of SU(2)}}
\put(46.6,-2.5){\makebox(2,1){b) Fundamental representation of SU(3) }}
\end{picture}\qquad\qquad
\includegraphics[scale=0.17]{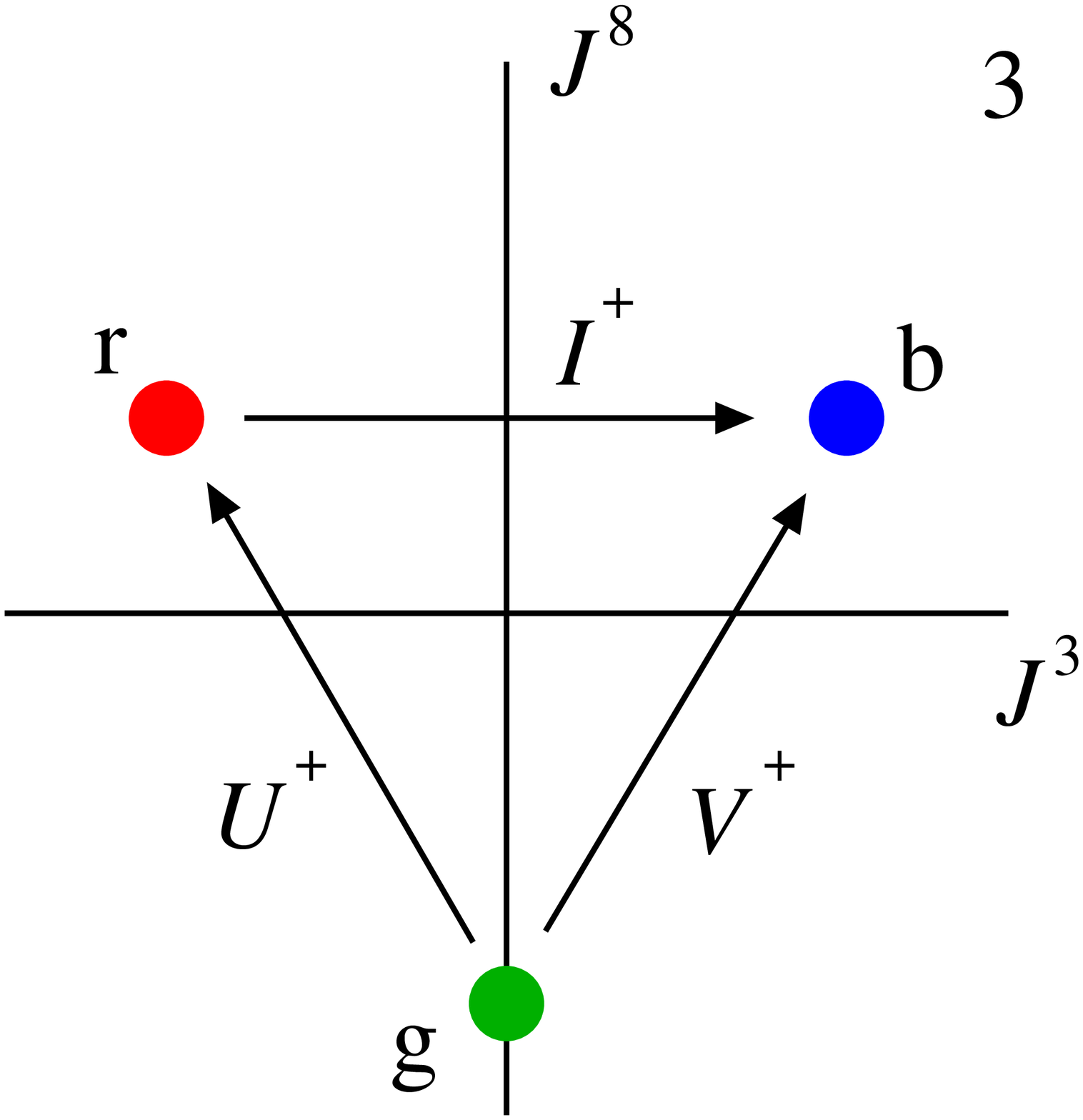}
\vspace{14pt}
\vchcaption{a) Weight diagram of the three-dimensional spin $S=1$
  representation of SU(2). The SU(2) weight diagrams are
  one--dimensional since there is only one diagonal generator ($S^z$)
  in the group SU(2). There is only one raising operator ($S^+$) and
  one lowering operator ($S^-$) and, hence, there is no generator
  connecting the $S^z=-1$ and the $S^Z=+1$ state directly. b) Weight
  diagram of the three--dimensional, fundamental representation of
  SU(3). SU(3) weight diagrams are two--dimensional since both $J^3$
  and $J^8$ are diagonal generators of the group SU(3). Due to the
  higher dimensionality of SU(3), each point in the weight diagram is
  directly connected with each other by the raising and lowering
  operators $I^+$, $I^-$, $U^+$, $U^-$, $V^+$, and $V^-$, as
  illustrated in the diagram.}
  \label{fig:su3-vs-spin1}
\end{vchfigure}
In the not-to distant future, it might be possible to realize SU(3)
spin chains in optical lattices of ultra-cold
atoms~\cite{greiter-07prb184441}.  There, one important challenge of
the implementation is that the system of the distinct atomic states
corresponding to blue, red, and green must be tuned in such a way that
the pairwise transition weights are equal, as the system only then
correctly resembles the SU(3) spin algebra with six raising/lowering
operators $I^{+}$, $I^{-}$, $U^{+}$, $U^{-}$, $V^{+}$, and $V^{-}$
linking the states of the fundamental representation with each other
(Fig~\ref{fig:su3-vs-spin1}b).  From this perspective, it becomes
immediately obvious that this is completely different from a
three-dimensional $S=1$ representation of SU(2), where the raising and
lowering operators $S^+$ and $S^-$ map the $S^z=0$ state to $S^z=\pm
1$, while there is no direct transition from $S^z=-1$ to $S^z=+1$ and
vice versa (Fig.~\ref{fig:su3-vs-spin1}a). It has also been proposed
that for approximately implemented SU($N$) chains of $N$-component
fermions, a molecular superfluid phase may
appear~\cite{capponi-08pra013624}.  Recently, SU(4) spin chains
attracted experimental interest in the field of transition-metal and
rare earth compounds, where such models appear to capture the physics
of coupled electronic and orbital degrees of
freedom~\cite{li-99prb12781,gu-02prb092404}.

From a field theoretical point of view, conformal field theories
(CFTs) have been enormously successful in describing the low energy
behavior of critical SU($N$) spin
chains~\cite{affleck-87prb5291,affleck88npb582,affleck86npb409}. In
this framework, critical means that the spins of the chain have a
diverging correlation length related to a gapless spectrum,
corresponding to the scale or, more precisely, conformal invariance of
the effective field theory describing these
systems~\cite{DiFrancescoMathieuSenechal97}.  The Wess-Zumino-Witten
model (WZW) plays a crucial role among those
models~\cite{wess-71plb95,witten84cmp455}.

Criticality is intimately related to the integrability of the SU($N$)
spin chain models which we study in this article. For the fundamental
representations of SU(2), SU(3), and SU(4), as well as higher
representations of SU(2) and SU(3), we numerically investigate
integrable models and their related CFTs, \ie the SU($N$) WZW models
of different levels $k$.  The Density Matrix Renormalization Group
(DMRG) method provides us with a highly suitable numerical method to
study the low energy sector of spin
chains~\cite{white92prl2863,schollwock05rmp259,noack-05proc,hallberg06advp477}.
From the results of our DMRG studies, we extract the central charge
and scaling dimension.  These parameters specify the associated
effective field theory.  We thus endeavor to establish numerically the
correspondence between CFTs and SU($N$) spin chains.

This article is organized as follows. In Section~\ref{sec:cft}, we
briefly review the basic features of CFT relevant to our
considerations, \ie the scaling dimension and the central charge. The
DMRG approach to SU($N$) spin chains is discussed in
Section~\ref{sec:dmrg}. In particular, we explain the implementation
of the SU($N$) spin algebra with its $N^2-1$ generators and discuss
the problem of convergence as $N$ is increased.  In
Section~\ref{sec:mod}, we introduce the integrable models we consider.
These include the nearest neighbor Heisenberg Hamiltonian for the
fundamental representations of SU(2), SU(3), and SU(4), as well as the
integrable Takhtajan--Babudjan Hamiltonians for $S=1$ and
$S=3/2$~\cite{takhtajan82pl479,babudjan82pl479,babudjan83npb317}.  We
also perform DMRG studies of the integrable SU(3) model with spins
transforming under the higher representation $\bs{6}$ proposed by
Andrei and Johanneson~\cite{andrei-84pl370,johannesson85npb235}.
The numerical results are presented in Section~\ref{sec:num}, and used
to extract the central charge and scaling dimension of the
corresponding WZW models.  
In Section~\ref{sec:con}, we conclude that the DMRG method can be
successfully applied to study SU($N$) spin chains.

\section{Conformal field theory, the central charge and scaling
  invariance}
\label{sec:cft}
The SU($N$) Wess--Zumino--Witten (WZW) models have been found to capture
the low energy behavior of a family of critical quantum spin
chains~\cite{affleck-87prb5291}.  WZW models are conformal field
theories, meaning that the Lagrangians are invariant under conformal
mappings. These are all combinations of translation, rotation, and
dilatation in two--dimensional space-time.  For field theories
with conformal invariance, it suffices to specify the scaling of
the fields or rather the scaling of their correlation functions to
characterize the theory
completely~\cite{DiFrancescoMathieuSenechal97}.  As such, once a CFT
is identified, there is no immediate need to work with the
associated Lagrangian.
Our emphasis in this article will be on the relation between the
universal parameters of the CFT and numerically accessible measures,
which we extract from the DMRG studies of the corresponding spin
chain models.  As a general structure, a WZW model consists of a
non-linear sigma model term and $k$ times a topological Wess-Zumino
term, where $k$ is a non-zero positive
integer~\cite{DiFrancescoMathieuSenechal97}.  The SU($N$) WZW model of
level $k$ (denoted SU($N)_k$ WZW in the following) can be
characterized by the central charge and the scaling dimension of the
primary field, both of which we will evaluate numerically in
Section~\ref{sec:num} below.  In the following formulas, subleading
finite size contributions are neglected if they appear.

\subsection{Central charge}
The central charge $c$ is defined in the framework of the Virasoro
algebra of the CFT~\cite{DiFrancescoMathieuSenechal97}.
Alternatively, $c$ is also named conformal anomaly number.  It appears
in the correlation function of the energy momentum tensor $T(z)$ of the
theory, where $z$ denotes a complex space-time variable.  This
correlation has a singularity as $z\rightarrow 0$, with a prefactor
proportional to $c$, $\langle T(z) T(0) \rangle \sim
\frac{c/2}{z^4}$. For the SU($N$$)_k$ WZW, $c$ is given by
\begin{equation}
  c=\frac{k(N^2-1)}{k+N}.
\label{ccharge}
\end{equation}
The for our purposes relevant feature of the central charge is that
$c$ appears as a universal scaling factor in the microscopically
accessible entanglement entropy~\cite{korepin92prl096402}.
%
Let $i$ denote a site, $L$ the total length of the spin chain,
$i=1,\dots, L$, and $\rho_\alpha$ the reduced density matrix where all
the degrees of freedom on sites $i>\alpha$ are traced out, \ie
$\rho_\alpha=\text{Tr}_{i>\alpha} \rho$. For this case, the
entanglement entropy is given by
\begin{equation}
  S_{\alpha,L}=-\text{Tr}\big[\rho_\alpha \log \rho_\alpha\big].
\label{ee}
\end{equation}  
For periodic boundary conditions and central charge $c$, the entropy
then takes the form~\cite{calabrese-04jsm06002}
\begin{equation}
  S_{\alpha,L}=\frac{c}{3} \log \left[\left(\frac{L}{\pi }\right)
    \sin{\left(\frac{\pi \alpha }{L}\right)}\right] 
  + c_1, 
\label{cc}
\end{equation}
where $c_1$ is a non-universal constant and the lattice spacing is set
to unity.  Thus, with $L$ being the total number of sites divided by
unit lattice spacing, the entanglement entropy obeys the symmetry
relation $S_{\alpha,L}=S_{L-\alpha,L}$ and has its maximum at
$\alpha=L/2$.  By virtue of~\eqref{cc}, $c$ can be extracted directly
from the entanglement entropy calculated via DMRG.

\subsection{Scaling dimension}
The scaling dimension $x$ is a property of the fields $\phi$
of the CFT~\cite{cardy84jpa385}. Conformal invariance implies that the
two-point correlation function of the field must satisfy
\begin{equation}
\langle \phi(z_1) \phi(z_2) \rangle = \vert f'(z_1)\vert^x 
\vert f'(z_2)\vert^x \langle \phi(f(z_1)) \phi(f(z_2)) \rangle ,
\end{equation}
where we constrain the conformal mapping $f(z)$ to a dilatation, and
$f'(z)$ is its derivative at point $z$.  For the finite systems we
study numerically, we can use that 
the low energy spectrum, and hence the energies of the finite system,
can be classified by the associated CFTs.
The lowest excited states (labeled by $p$) above the ground state
($p=0$) belong spectrally to a conformal tower~\cite{affleck-89jpa511},
with energies which obey the relation
\begin{equation}
E_{p,L}-E_{0,L}=\frac{2\pi v}{L} x_p,
\label{x}
\end{equation}
where $x_p$ denotes the scaling dimension of the field associated with
the $p$th state, and $v$ is the Fermi velocity.  The dependence on $v$
reflects that in the low energy limit, the only relevant momentum
scale of the spin chain is provided by the linearized dispersion
around the Fermi points.  This allows us to extract the scaling dimension
times the Fermi velocity, $x_pv$, as the energies in the l.h.s.\
of~\eqref{x} are numerically accessible through DMRG.  In the
following, we shall focus on the first excited state $x_1 \equiv x$,
\ie the scaling dimension of the primary field.

\subsection{Fermi velocity parameter}
In view of~\eqref{cc} and~\eqref{x}, it is clear that we need one
further relation, to extract the Fermi velocity $v$ from our numerical
studies.  The required relation is 
\begin{equation}
E_{0,L}=E_{0,\infty}-\frac{\pi c v}{6L},
\label{ediff}
\end{equation}
where $E_{0,L}$ and $E_{0,\infty}$ denote the ground state energies of
the finite and the infinite chain, respectively.  This relation can be
easily understood from the field theoretical point of
view~\cite{affleck-89jpa511}: For a finite length and temperature
$T=0$, $L$ sets the inverse energy scale of the system.  This scale
can be rephrased in terms of a field theory at finite temperature and
no length scale, \ie an infinite chain at temperature $T=v/L$.

Writing~\eqref{ediff} in terms of the free energy density for this
finite temperature field theory, we obtain the correct specific heat
linear in $T$, as we expect for a gapless spectrum.  As we calculate
$E_{0,L}$ directly and extract $e_{0,\infty}=E_{0,L}/L$ for
$L\to\infty$ by finite size scaling, we can obtain $v$ from
\eqref{ediff} once we have obtained $c$ from~\eqref{cc}.

\section{The DMRG method}
\label{sec:dmrg}

\begin{vchfigure}[h]
  \centering
  \includegraphics[scale=0.9]{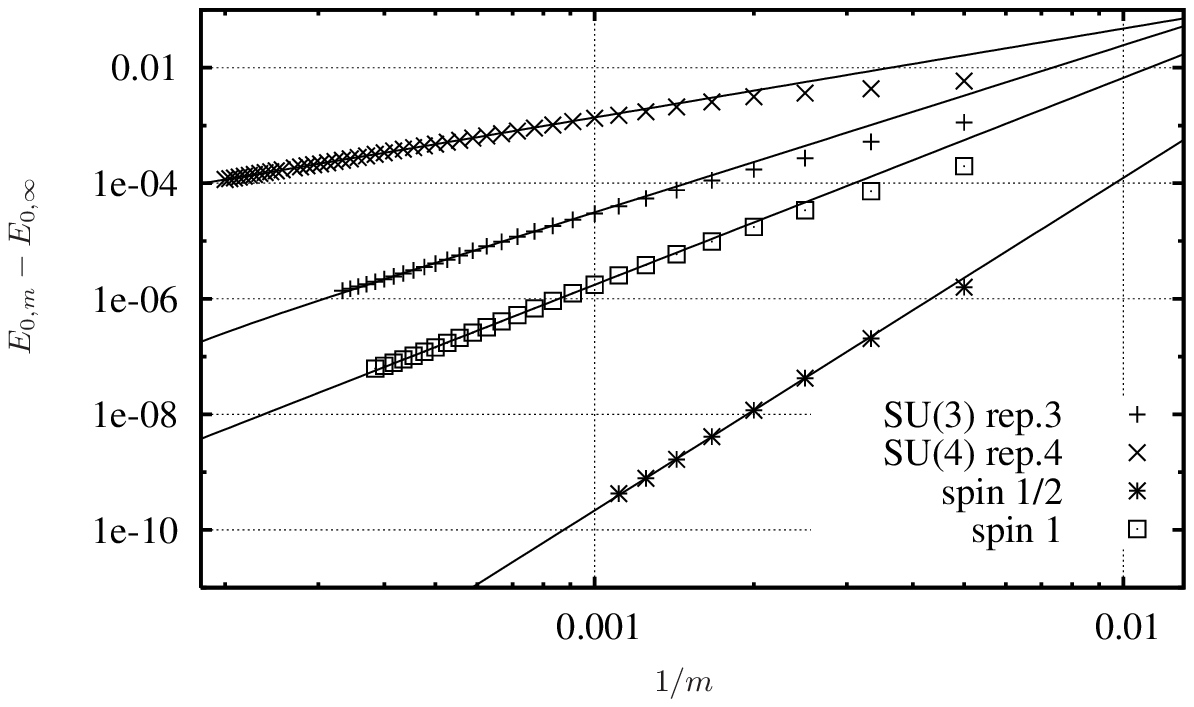}
  \vchcaption{Periodic Boundary Conditions: logarithmic plot for the
    energy difference of the finite size ground state energy and the
    thermodynamic site limit $L\rightarrow \infty$ versus inverse
    number of states $m$ kept in the DMRG sweeps. Shown are the lines
    for the nearest neighbor Heisenberg model in the fundamental
    representations of SU(2), SU(3), and SU(4), as well as for the
    $S=1$ TB model for comparison.  The length of the chain is $L=48$.}
  \label{fig:convergence-PBC}
\end{vchfigure}

In the last decade, the DMRG was successfully applied to numerous
SU(2) spin models. Very recently, the DMRG was further used to
investigate the
SU(3) representation $\bs{3}$ Heisenberg model~\cite{corboz-07prb220404}.
Here we generalize
to the six--dimensional representation $(2,0)\equiv\bs{6}$ of SU(3),
which is formed by symmetric combination of two fundamental
representations of SU(3).  We also present DMRG studies of a spin
chain with spins transforming under the fundamental representation of
SU(4).  Our work hence requires the explicit implementation of the
su(3) and su(4) spin algebras with its $N^2-1$ generators, which are
explicitly given in Apps.~\ref{app:SU3R6matrices}
and~\ref{app:SU4R4_matrices}.  Note that this implementation is more
involved than the implementation of SU($N$) Hubbard models, where the
explicit spin algebra does not enter, and the SU($N$) symmetry enters
only through the number of different fermionic species.

\begin{vchfigure}[t]
  \centering
  \includegraphics[scale=0.9]{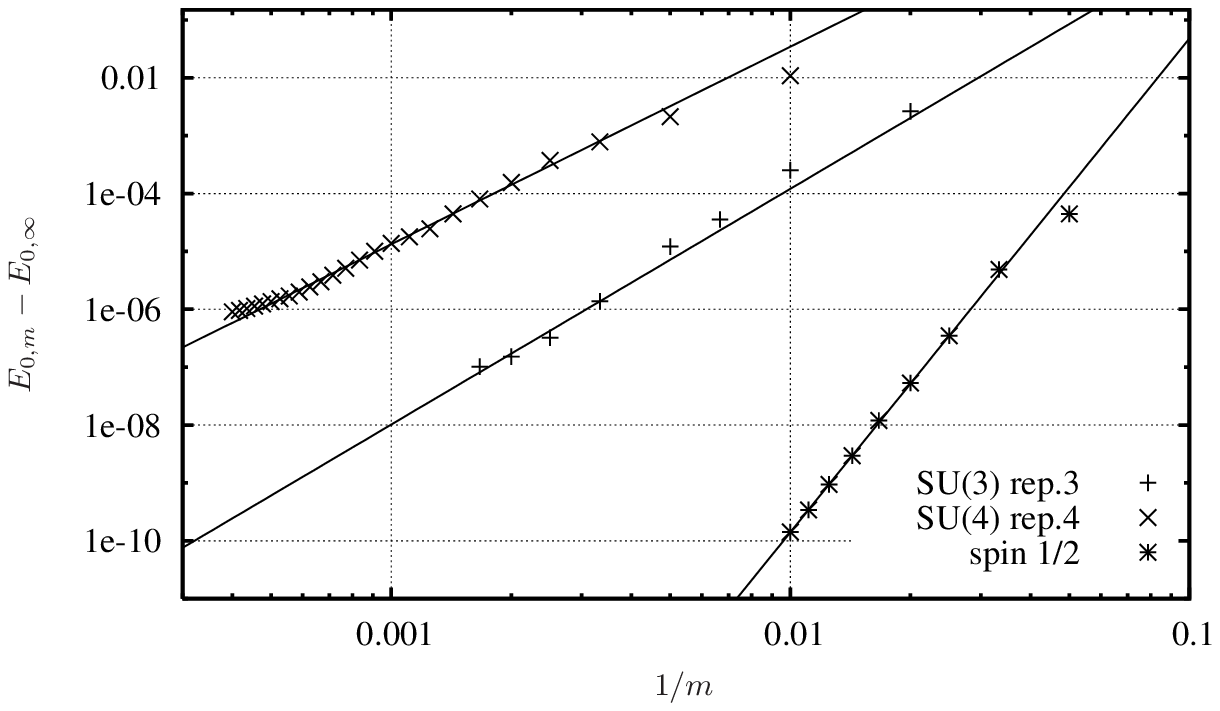}
  \vchcaption{Hard Wall Boundary Conditions: logarithmic plot for the
    energy difference of the finite size ground state energy and the
    thermodynamic limit $L\rightarrow \infty$ versus inverse number of
    states $m$ kept in the DMRG sweeps. Shown are the lines for the
    fundamental representations of SU(2), SU(3), and SU(4).  The
    length of the chain is again $L=48$.  As compared to the case of
    PBCs shown in Fig.~\ref{fig:convergence-PBC}, the system converges
    much faster for a comparable number of states kept in the DMRG
    sweeps.}
  \label{fig:convergence-OBC}
\end{vchfigure}

An important problem with numerical studies of SU($N$) spin chains in
general is the increasing dimensionality of the subspace of one site
of the chain, due to either larger values for $N$ or higher spin
representations (like rep $\bf 6$ for SU(3)).  For DMRG, of course,
this dimensionality limits the system sizes we can access.  In
Fig.~\ref{fig:convergence-PBC}, we have plotted the convergence of the
DMRG iteration as the number $m$ of states kept in the effective
density matrix is increased. We observe that for comparable $m$, the
convergence decreases rapidly as we go to higher SU($N$), according
for the exponential increase of the Hilbert space as the number of
states per site grows.  However, as confirmed by our numerical results
reported below, our DMRG code is capable of at least handling critical
spin chains up to SU(4) with reasonable convergence and accuracy.
While the plots in Fig.~\ref{fig:convergence-PBC} are obtained using
periodic boundary conditions (PBCs), the convergence behavior for hard
wall boundary conditions (HWBCs) is shown in
Fig.~\ref{fig:convergence-OBC}.  Note that HWBCs rather than PBCs are
the natural choice for DMRG, as the number of DMRG states $m$ we need
to keep to achieve a similar level of precision for PBCs is, according
to our calculations, roughly the square of the number of states we
need to keep for HWBCs.  Nonetheless, the results we present below are
obtained with PBCs, as PBCs allow a more convenient treatment of the
finite size corrections for the quantities we extract.  In particular,
\eqref{ediff} is valid only for PBCs.  (There is a relation
corresponding to~\eqref{cc} for HWBCs~\cite{calabrese-04jsm06002}.)
We have used 10 DMRG sweeps for all the calculations we present.

\begin{vchfigure}[h]
  \centering
  \includegraphics[scale=1.0]{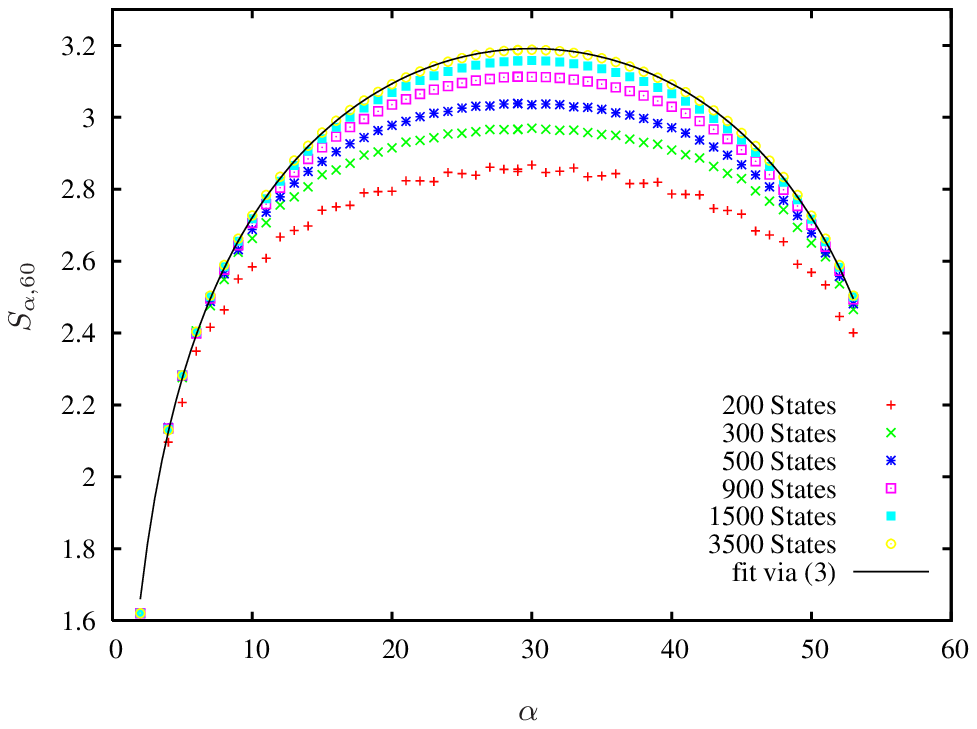}
  \vchcaption{The entanglement entropy $S_{\alpha,60}$ as another example
    for the convergence behavior of the SU(3) Heisenberg model. (a)
    shows the EE $S_{\alpha,60}$.  The different curves correspond to
    different number of kept DMRG states (200, 300, 500, 900, 1500,
    3500 states). The system with 3500 kept DMRG states is fully
    converged and provides a benchmark. The truncated Hilbert space for this converged job contains about 8 million states.}
  \label{fig:ee-comparison-su3}
\end{vchfigure}
As a demonstration of convergence, Fig.~\ref{fig:ee-comparison-su3}
shows the entanglement entropy for different numbers of kept DMRG
states $m$ for the SU(3) representation ${\bf 3}$ Heisenberg model.  The result
displays the $S_{\alpha,L}=S_{L-\alpha,L}$ symmetry mentioned above
and fits the prediction \eqref{cc} to astonishing accuracy.  We find,
however, that this accuracy requires a number $m$ of states kept which
is large in comparison with standard applications of the DMRG method,
and which demands large computational resources.  This is partially
due to the criticality of the models we study.  With a spectrum that
is gapless in the thermodynamic limit, a large subspace of the entire
Hilbert space contributes to the long range correlations, which is
reflected in a large number of relevant weights in the density matrix.
Nonetheless, with a sufficiently high value of states kept, very
accurate results can be extracted from the DMRG computations.  Even
for rather small systems consisting of $\mathcal{O}$(100) sites, we
obtain highly accurate estimates for the central charges of the
critical models described in the following section.  As the
entanglement entropy is not directly accessible by other numerical
methods, 
the DMRG method is preeminent to our purposes.

\section{Integrable models of critical SU($\bs N$) chains}
\label{sec:mod}

The SU($N$) spin chain models we investigate numerically in this work
are described by a family of Hamiltonians $\mathcal{H}^{[N,m]}$, which
are amenable to the transfer matrix
method~\cite{takhtajan82pl479,babudjan82pl479,babudjan83npb317,
andrei-84pl370,johannesson85npb235,sutherland75prb3795}.
Note that some of the models $\mathcal{H}^{[N,m]}$ were investigated
by numerical and analytical solutions of the Bethe ansatz
equations~\cite{alcaraz-88jpa4397,alcaraz-89jpal865,martins90prl2091}.
The representations $[N,m]$ of SU($N$) are given by the totally 
symmetric combination of $m$ fundamental representations of SU($N$).
The corresponding Young tableaux is
\begin{displaymath}
    \setlength{\unitlength}{8pt}
    \begin{picture}(5,5)(0,0.15)
      \put(0,4){\line(1,0){5}}
      \put(0,3){\line(1,0){5}}
      \put(0,3){\line(0,1){1}}
      \put(1,3){\line(0,1){1}}
      \put(2,3){\line(0,1){1}}
      \multiput(2.7,3.48)(0.3,0){3}{\circle*{0.1}} 
      \put(4,3){\line(0,1){1}}
      \put(5,3){\line(0,1){1}}
      \put(2.5,1.5){\makebox(0,0)%
        {$\underbrace{\quad\text{~~~}\qquad}_{\text{\small $m$~boxes}}$}}
      \put(5.8,3.06){\circle*{0.1}} 
   \end{picture} 
\end{displaymath}
For SU(2), all the representations are of this form, with $m=2S$.  For
SU(3), the symmetric representations include the fundamental
representation $\bf 3$ and the representation $\bf 6$, for $m=1$ and
$2$, respectively.  The dimensionality $n$ of the totally symmetric
representation $[N,m]$ is in general given by
\begin{equation}
  n\,\equiv\, {\rm dim}[N,m]\,=\,{N-1+m\choose m}.
\end{equation}
The Hamiltonians $\mathcal{H}^{[N,m]}$ contain two-site interactions
only and are invariant under global SU($N$) spin rotations
, \ie Heisenberg interaction terms to arbitrary
power~\cite{takhtajan82pl479,babudjan82pl479,babudjan83npb317,
andrei-84pl370,johannesson85npb235,sutherland75prb3795}.
Note that all the models $\mathcal{H}^{[N,m]}$ are integrable, due to
an infinite number of operators which commute with the Hamiltonians.
In this work, we consider the models with $[N,m]=[2,1]$, $[2,2]$,
$[2,3]$, $[3,1]$, $[3,2]$, and $[4,1]$. 

The Hamiltonians for $[N,1]$, \ie the fundamental representations, are
just the nearest-neighbor Heisenberg models,
\begin{equation}
  \label{nnHM-SU(N)}
  \mathcal{H}^{[N,1]}=\sum_{i=1}^{\mathcal{N}} \bs{S}_i\bs{S}_{i+1}.
\end{equation}
In general, $\bs{S}_i$ is an SU($N$) representation $[N,m]$ spin
operator at site $i$.  Since the dimension of the Lie algebra su($N$)
is $N^2-1$, the spin operator $\bs{S}_i$ consists of the $N^2-1$
generators,
\begin{equation}
  \label{sun-spin-operators}
  S_i^{\alpha} = \frac{1}{2} \sum_{\sigma,\sigma'=f_1,\ldots,f_n}
c_{i\sigma}^{\dagger}V^{\alpha}_{\sigma\sigma'}c_{i\sigma'}^{\phantom{\dagger}},
\end{equation}
where $\alpha=1,\ldots,N^2-1$, $V^{\alpha}_{\sigma\sigma'}$ are the
SU($N$) Gell-Mann matrices, and $f_1,\ldots,f_n$ denote the $n$
different spin states~\cite{Cornwell84vol2}.    Trivially,
$\bs{S}_i\bs{S}_{i+1}\equiv\sum_{\alpha=1}^{N^2-1} S_i^{\alpha}
S_{i+1}^{\alpha}$.

For the fundamental representation $[2,1]$ of SU(2), the $V$'s are
just the Pauli matrices and the two spin states can be classified by
the eigenstates $f_1=\uparrow$, $f_2=\downarrow$ of $S^z$.  For the
fundamental representation $[3,1]$ of SU(3), the $V$'s are given by
the eight Gell-Mann matrices.  The matrices $V$ for representations
$[3,2]$ (\ie SU(3) representation $\bs{6}$) and $[4,1]$ (\ie SU(4)
representation $\bs{4}$) are written out in
Apps.~\ref{app:SU3R6matrices} and~\ref{app:SU4R4_matrices}.  In our
numerical implementations, we have scaled the Hamiltonians such that
the pre-factor of the bilinear Heisenberg term $\bs{S}_i\bs{S}_{i+1}$
is $\pm 1$ and we have dropped the constant term.

As we confirm numerically below, the low-energy behavior of the
models $\mathcal{H}^{[N,1]}$ is described by the SU($N$)$_1$ WZW
model, with topological coupling constant $k=1$.
With~\eqref{ccharge}, we expect to find $c=N-1$ for the central
charge.  The integrable spin $S=1$ model we investigate, the
Takhtajan--Babudjan model~\cite{takhtajan82pl479,
  babudjan82pl479,babudjan83npb317}, is given by
\begin{equation}
  \label{TB-model}
  \mathcal{H}^{[2,2]}=\sum_{i=1}^{\mathcal{N}} \left[ \bs{S}_i\bs{S}_{i+1}
-\left( \bs{S}_i\bs{S}_{i+1} \right)^2 \right].
\end{equation}
The low energy physics is described by the SU(2)$_2$ WZW model.
With~\eqref{ccharge}, we expect to find $c=\frac{3}{2}$.  Note that
the criticality of this integer spin model is not inconsistent with
the Haldane gap, as Haldane's classification applies to {\it generic}
integer spin chains, while the Takhtajan--Babudjan
model~\eqref{TB-model} is tuned to criticality.

The next higher dimensional integrable SU(2) model from the
Takhtajan--Babudjan series is given by the spin 3/2 Hamiltonian
\begin{equation}
  \label{S=3/2-model}
  \mathcal{H}^{[2,3]}=\sum_{i=1}^{\mathcal{N}} \left[ -\bs{S}_i\bs{S}_{i+1}
  +\frac{8}{27}\left( \bs{S}_i\bs{S}_{i+1} \right)^2 
  +\frac{16}{27}\left( \bs{S}_i\bs{S}_{i+1} \right)^3 \right].
\end{equation}
The corresponding CFT is the SU(2)$_3$ WZW model, which implies that the 
central charge is $c=\frac{9}{5}$.

Finally, the Andrei--Johannesson~\cite{andrei-84pl370,johannesson85npb235} 
model consists of SU(3) spins transforming under the six--dimensional 
representation $\bs{6}$, and is given by
\begin{equation}
  \label{su3rep6-model}
  \mathcal{H}^{[3,2]}=\sum_{i=1}^{\mathcal{N}} \left[ \bs{S}_i\bs{S}_{i+1}
-\frac{3}{5}\left( \bs{S}_i\bs{S}_{i+1} \right)^2 \right].
\end{equation}
The corresponding CFT is the SU(3)$_2$ WZW model, which implies 
$c=\frac{16}{5}$.  We now turn to our numerical results for these models.

\section{Numerical results}
\label{sec:num}
\subsection{Central charge}
\begin{vchfigure}[h!]
  \centering
  \includegraphics[scale=0.85]{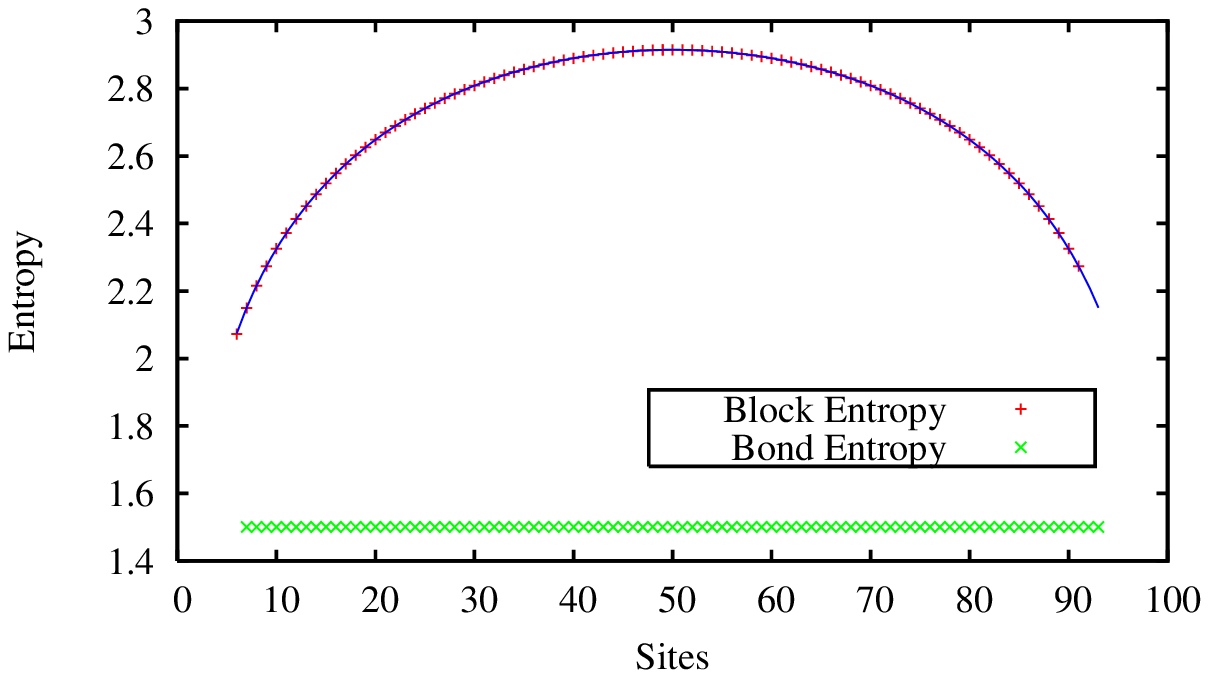}
  \vchcaption{Entanglement entropy (block entropy) of the integrable
    SU(2) $S=1$ Hamiltonian with PBCs. The solid line corresponds to
    the formula \eqref{cc} with the fit parameter $c$, the central
    charge of the corresponding CFT. The uniform bond entropy, \ie the
    nearest neighbor entanglement entropy, indicates a homogeneous and
    translationally invariant ground state.}
  \label{fig:ee-s2-100-4000}
\end{vchfigure}
\begin{vchfigure}[h!]
  \centering
  \includegraphics[scale=0.9]{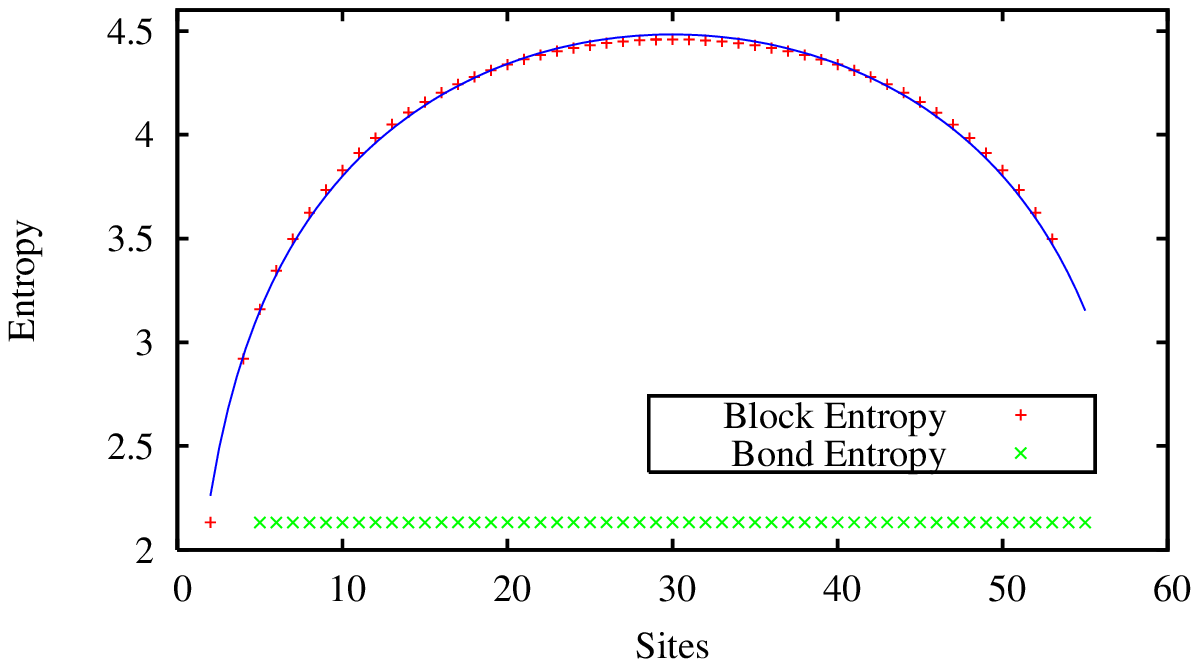}
  \vchcaption{Entanglement entropy (block entropy) of the SU(4)
    nearest neighbor Heisenberg model with PBCs.  The solid line
    constitutes a fit of the data using~\eqref{cc}, yielding a central
    charge close to the predicted value $c=3$ (for details see
    Tab.\,\ref{tab:num-results}).}
  \label{fig:ee-su4-60-8000}
\end{vchfigure}

As noted above, the entanglement entropy is provided quite naturally
in DMRG, as in each sweep we really calculate reduced density
matrices, from which we easily obtain the entanglement entropy
via~\eqref{ee}.  From the plots of the entanglement entropy vs.\ the
site index, we obtain a numerical value for the central charge
via~\eqref{cc}.  In Fig.~\ref{fig:ee-s2-100-4000}, we show the
entanglement entropy (also called block entropy) for the integrable
$S=1$ Hamiltonian, the Takhtajan--Babudjan model, as an illustrative
example. The fit yields a central charge of $c=1.50717 \pm 0.0003$,
where the error corresponds to the fitting error shown in
Tab.~\ref{tab:num-results}. The result is in excellent agreement with
the value $c=\frac{3}{2}$ predicted by CFT.  We have also plotted the
bond entropy, which is the entanglement entropy of two neighboring
sites $\alpha$ and $\alpha+1$ with the remainder of the system.  In
general, a bond entropy which is not site independent indicates a
spontaneous breakdown of translational invariance (like \eg
dimerization) in the ground state.
Despite being a quantity of its own interest to extract information
from finite systems~\cite{molina-07prb235104}, we attach no
significance to the bond entropy beyond the confirmation of
translational invariance.
Other quantities, \eg the ground state
stiffness~\cite{schmitteckert-04prb195115}, could in principle be
studied within DMRG to supplement the ground state studies, but are
not our point of consideration in this work.

\begin{vchtable}[t]
  \vchcaption{Theoretical predictions and numerical results for the
    central charge for the Hamiltonians we considered.  The error
    quoted are due to inaccuracies when fitting the data for
    entanglement entropy obtained numerically to~\eqref{cc}.  An
    additional systematic error, which we have not estimated
    separately, arises from the states discarded within the DMRG.}
\label{tab:num-results}\renewcommand{\arraystretch}{1.5}
\begin{tabular}{cccccc} \hline
Hamiltonian $[N,m]$ & $N$ & $k$ & $c$ & $c_{\rm DMRG}$ \\
\hline
$[2,1]$ & 2 & 1 & 1 & $1.0001 \pm 0.0012$  \\
$[2,2]$ & 2 & 2 & $\frac{3}{2}$ & $1.5072 \pm 0.0003$  \\
$[2,3]$ & 2 & 3 & $\frac{9}{5}$ & $1.8002 \pm 0.0211$   \\
$[3,1]$ & 3 & 1 & 2 & $2.0001 \pm 0.0102$   \\
$[3,2]$ & 3 & 2 & $\frac{16}{5}$ & $3.2214 \pm 0.0437$ \\
$[4,1]$ & 4 & 1 & 3 & $2.9527 \pm 0.0237$   \\
\hline
\end{tabular}
\end{vchtable}
The discrepancy between the data and the fit to~\eqref{cc} is most
visible at the maximum of the entanglement entropy, \ie for half of
the sites traced out, as this discrepancy is due to entropy we have
discarded by discarding states.  While there is no difference visible
in Fig.~\ref{fig:ee-s2-100-4000}, a discrepancy can be discerned in
Fig.~\ref{fig:ee-su4-60-8000}, where we have plotted the entanglement
entropy of the SU(4) Heisenberg model in the fundamental
representation.  This small discrepancy is present even though we keep
8000 DMRG states for the sweep iterations which results in a truncated
Hilbert space containing 22 million states. Note that the DMRG
calculation for the ground state of this model with 8000 kept states
has taken 68 hours of computer time (4 CPU cores) with ca.\,20 gigabyte
of memory, while the same calculation with 5500 kept states has taken
28 hours (4 CPU cores) with ca.\,9 gigabyte of memory. 
Here we only exploited the abelian quantum numbers of SU($N$). By
using the square of the total spin $\bs{S}^2$ as additional quantum
number and representing the states according to the Wigner--Eckhardt
theorem, one should be able to reduce the required Hilbert space
significantly. However, already for SU(2), the Clebsch--Gordon
coefficients make the implementation cumbersome~\cite{mcculloch}, and
for SU(3) and SU(4) the corresponding Clebsch--Gordon coefficients are
more complicated. Therefore we decided not to implement them. Note
that the use of non--abelian quantum numbers does not automatically
lead to better performance, \eg in SU(2) it helps for small $S$
sectors only, since otherwise the sum over reduced states gets too
involved.

The value we obtain for the central charge, $c=2.95268 \pm 0.02368$,
however, is reasonably close to the predicted value $c=3$.  In
addition to the fitting error we quote in Tab.\,\ref{tab:num-results},
there is a systematic error due to the entropy we have discarded by
discarding states in DMRG.  As all the contributions to the
entanglement entropy are positive definite, this systematic error
leads to a slight underestimate for the numerically obtained central
charge.
Our results for the models introduced in Section~\ref{sec:mod} are
presented in Tab.~\ref{tab:num-results}. In general, we find excellent
agreement between analytical values and numerical data.


\subsection{Scaling dimension of critical models}
\begin{vchfigure}[b]
\includegraphics[scale=0.8]{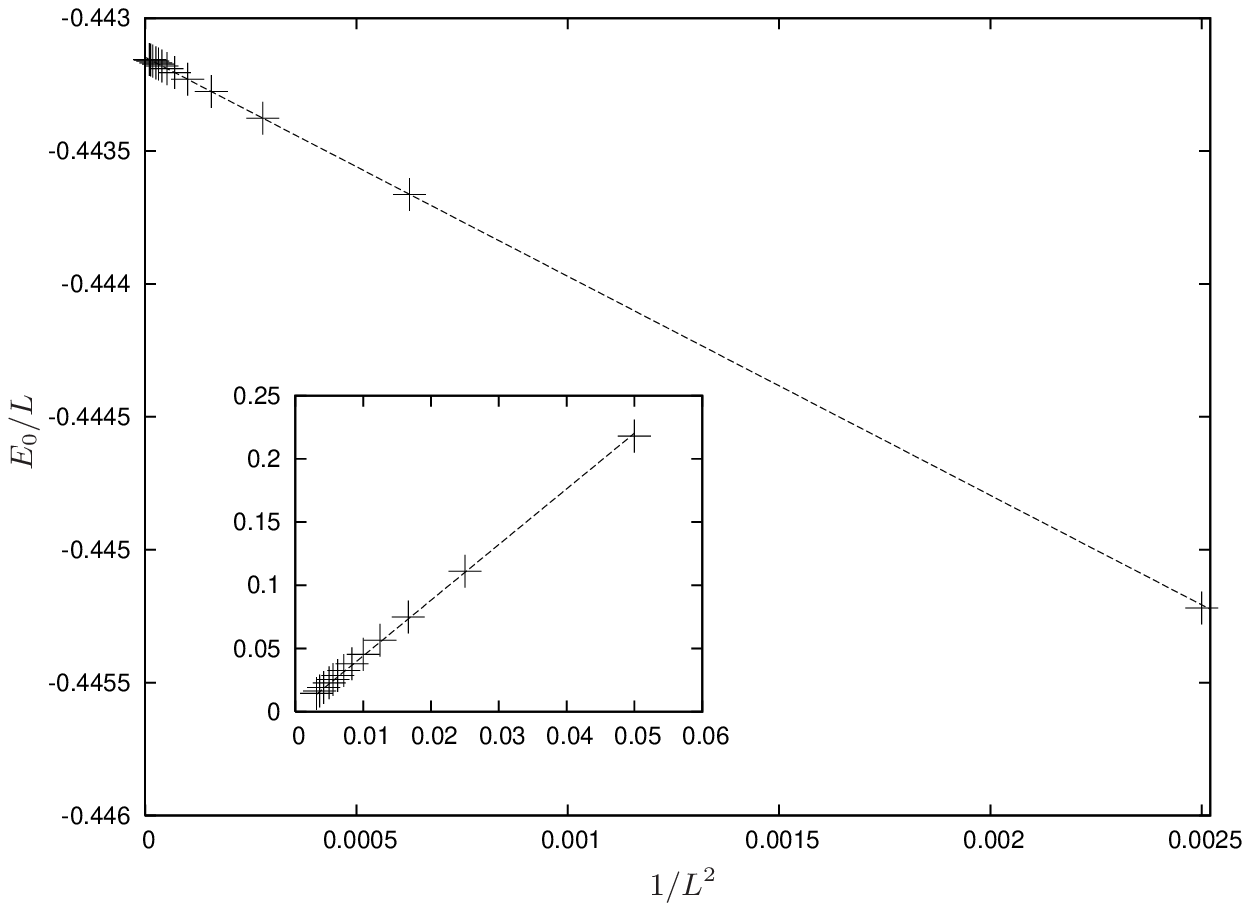}
\vchcaption{Scaling dimension for the spin 1/2 Heisenberg model.
  According to Eq. \eqref{ediff} the Fermi velocity is fitted in the
  main picture ($v \sim \frac{\pi}{2}$), which is used as a parameter
  for the fit of the scaling dimension done in the inset. In the
  inset, $E_{1,L}-E_{0,L}$ is plotted vs. $1/L$ and according to Eq.
  \eqref{x} we have fitted the scaling dimension $x=0.443\pm 0.0020$.
  The data points correspond to chains (PBCs) from 20 to 320 sites.}
  \label{fig:spin1/2-sd}
\end{vchfigure}

\begin{vchtable}[t]
  \vchcaption{Theoretical predictions and numerical results for the
    scaling dimension of the primary fields for SU(2) and SU(3)
    Hamiltonians. The deviations from the analytical values are higher
    than the deviations for central charges discussed above. The errors
    quoted again refer to inaccuracies of the fits only.}
\label{tab:num-results-x}\renewcommand{\arraystretch}{1.5}
\begin{tabular}{cccccc} \hline
Hamiltonian $[N,m]$ & $N$ & $k$ & $x$ & $x_{\rm DMRG}$ \\
\hline
$[2,1]$ & 2 & 1 & $\frac{1}{2}$ & $0.443 \pm 0.0020$  \\
$[2,2]$ & 2 & 2 & $\frac{3}{8}$ & $0.338 \pm 0.0006$  \\
$[3,1]$ & 3 & 1 & $\frac{2}{3}$ & $0.638 \pm 0.0010$   \\
\hline  
\end{tabular}
\end{vchtable}
From the spectrum we numerically calculate via DMRG, it takes two
steps to obtain an estimate for the scaling dimension of the primary
field.  First, with the central charge obtained through the
entanglement entropy, we use \eqref{ediff} to arrive at an estimate
for the Fermi velocity $v$.  Second, we use~\eqref{x} to obtain an
approximate value for the scaling dimension $x$.  The values we obtain
for some of the models we study are listed in
Tab.~\ref{tab:num-results-x}.

Both steps require only linear fits, which are easily accomplished.
For the fundamental representations of SU(2) and SU(3), these fits are
shown in Figs.~\ref{fig:spin1/2-sd} and \ref{fig:su3-sd}.  In the
process, however, we omit significant finite size corrections.  To
begin with, in~\eqref{ediff}, marginal sub-leading contributions of
the order of $\mathcal{O}(\frac{1}{L (\log L)^3})$ are
omitted~\cite{cardy86jpa1093}.  As we consider spin chains with of the
order of 100 sites, these corrections are at the order of $1\%$ and
thus for our purposes negligible.  In~\eqref{x}, however, the error
due to omitting marginal contributions is of order
$\mathcal{O}(\frac{1}{L \log L})$, and hence significantly larger.
This error is essentially responsible for the discrepancy between the
analytical results and the numerical findings in
Tab.~\ref{tab:num-results-x}.  For these reasons, our numerical
results for the scaling dimension are not nearly as accurate as for
the central charges, where finite size corrections did not enter.  By
use of non-Abelian bosonization~\cite{affleck-89jpa511}, the
logarithmic correction can be calculated in principle.  For the case
$S=1$ of SU(2), this has been carried out by Hijii and
Nomura~\cite{hijii-02prb104413}.  For our purposes, however, such an
analysis is not required.  
\begin{vchfigure}[b]
\includegraphics[scale=0.8]{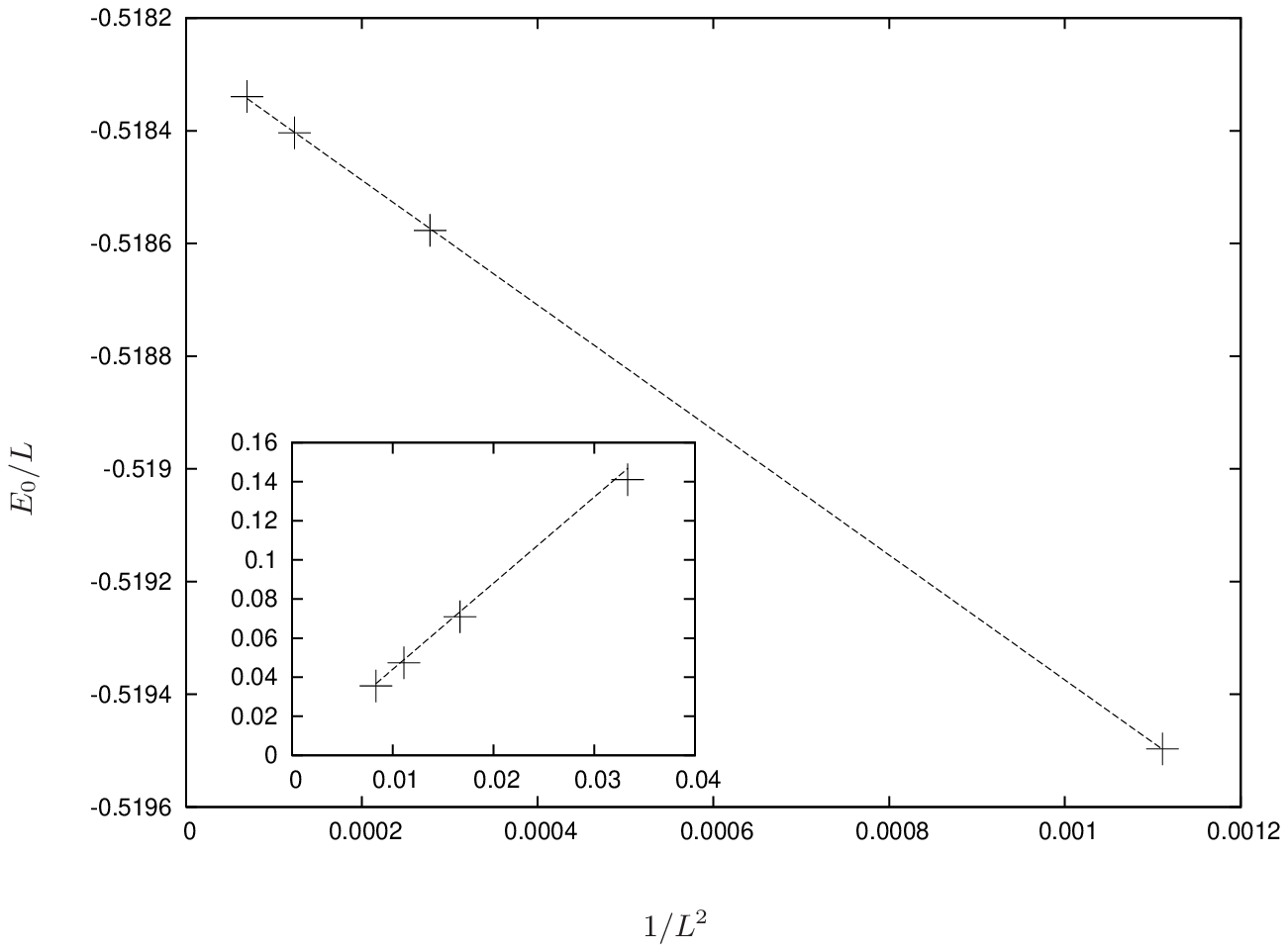}
\vchcaption{Scaling dimension for the SU(3) representation $\bs{3}$
  Heisenberg model. Less data points have been computed compared to
  $S=1/2$ SU(2) in Fig.~\ref{fig:spin1/2-sd}, but yields a similar
  numerical fit precision. The Fermi velocity is fitted to $v \sim
  \frac{\pi}{3}$. In the inset the energy difference $E_{1,L}-E_{0,L}$
  is plotted vs. $1/L$ and according to Eq. \eqref{x} we have fitted
  the scaling dimension $x=0.638\pm 0.0010$. The data points
  correspond to chains (PBCs) with 30, 60, 90, and 120 sites.}
  \label{fig:su3-sd}
\end{vchfigure}
The fits to leading order for the cases
$S=1/2$ and $S=1$ of SU(2) as well as the fundamental
representation~$\bs{3}$ of SU(3) are sufficiently conclusive to
identify the scaling dimension of the primary fields of the
corresponding WZW model.  For a more refined spectral analysis or
calculation of the various scaling dimensions of the descendants of
the primary fields, however, it would be indispensable to include the
marginal contributions into the fits as well.

\section{Conclusion}
\label{sec:con}
To summarize, we have investigated critical spin models of higher
representations of SU(2), SU(3) and SU(4) by DMRG, extracting the
central charge as well as the scaling dimension of the primary field
from our numerical results.  These results agree accurately with the
predictions of the associated conformal field theories, the
SU($N$)$_k$ WZW models.  We have thus shown that the study of block
entropies within DMRG is a suitable numerical tool to investigate
SU($N$) spin chains including higher representations.  It thus
represents a fruitful method to complement analytical approaches to
these models and perspectively provide important information on models
where analytical methods may not be practicable or even applicable.

\begin{acknowledgement}
  We thank D.~Schuricht for useful discussions and a critical reading
  of the manuscript. SR is supported by the Cusanuswerk, RT by the
  Studienstiftung des deutschen Volkes.  MG and PS are supported by the
  Deutsche Forschungsgemeinschaft (DFG) through grant FOR 960.
\end{acknowledgement}

\vspace{20pt}

\appendix

\section{Gell--Mann matrices for the fundamental 
representation {\bf 3} of SU(3) }

The algebra su(3) has two diagonal generators $V^3$ and $V^8$.  The
SU(3) Gell-Mann matrices for the fundamental representation are given
by~\cite{Cornwell84vol2}\\
\begin{displaymath}
\begin{array}{c@{}c@{}c@{\hspace{6pt}}c@{}c@{}c@{\hspace{6pt}}c@{}c@{}c}
V^1&=&\!\left(\begin{array}{ccc}0&1&0\\1&0&0\\0&0&0\end{array}
\right)\quad&
V^2&=&\!\left(\begin{array}{ccc}0&-i&0\\i&0&0\\0&0&0\end{array}
\right)\quad&
V^3&=&\!\left(\begin{array}{ccc}1&0&0\\0&-1&0\\0&0&0\end{array}
\right)\\[27pt]
V^4&=&\!\left(\begin{array}{ccc}0&0&1\\0&0&0\\1&0&0\end{array}
\right)\quad&
V^5&=&\!\left(\begin{array}{ccc}0&0&-i\\0&0&0\\i&0&0\end{array}
\right)\quad&
V^6&=&\!\left(\begin{array}{ccc}0&0&0\\0&0&1\\0&1&0\end{array}
\right)\\[27pt]
V^7&=&\!\left(\begin{array}{ccc}0&0&0\\0&0&-i\\0&i&0\end{array}
\right)\quad&
V^8&=&
\multicolumn{3}{l}{{\displaystyle\frac{1}{\sqrt{3}}}
\!\left(\begin{array}{ccc}1&0&0\\0&1&0\\0&0&-2\end{array}\right)\!\!.}
\end{array}
\end{displaymath}
They are normalized as
$\mathrm{tr}\left(V^aV^b\right)=2\delta_{ab}$ and satisfy
the commutation relations
$\comm{V^a}{V^b}=2f^{abc}V^c.$ The structure
constants $f^{abc}$ are totally antisymmetric and obey Jacobi's
identity 
\begin{displaymath}
f^{abc}f^{cde}+f^{bdc}f^{cae}+f^{dac}f^{cbe}=0.
\end{displaymath} 
Explicitly, the non-vanishing structure constants are given by $f^{123}=i$,
$f^{147}=f^{246}=f^{257}=f^{345}=-f^{156}=-f^{367}=i/2$,
$f^{458}=f^{678}=i\sqrt{3}/2$, and 45 others obtained by permutations of 
the indices.

\section{Matrices for the representation {\bf 6} of SU(3)}
\label{app:SU3R6matrices}
For completeness, we write out the matrix representation of the SU(3)
generators for representation $\bs{6}$, as those are rarely given explicitly
in the literature.  As illustrated in Fig~\ref{fig:su3-vs-spin1}b (or
also from the Gell--Mann matrices above),
su(3) possesses two diagonal generators $J^3$ and $J^8$ and 6 ladder
operators, where always pairs like $I^+, I^-$ are adjoint counterparts
or hermitian conjugates of each other.  In the notation of Eq.
\eqref{sun-spin-operators}, $J^3_{ \scriptstyle{ \textrm{ rep.6 }} }$
corresponds to $S^3$ and likewise $J^8_{ \scriptstyle{ \textrm{ rep.6
    }} } \equiv S^8$. The ladder operators are connected to the spin
operators by $S^1 \pm i S^2 = I^{\pm}$, $S^4 \pm i S^5 = U^{\pm}$, and
$S^6 \pm i S^7 = V^{\pm}$.  In the following, dots denote zeroes:
\begin{longtable}{LL}
J^3_{ \scriptstyle{ \textrm{ rep.6 }} } = \left( \begin{array}[c]{cccccc}
 1&\cdot&\cdot&\cdot&\cdot&\cdot\\[4pt] 
 \cdot&\cdot&\cdot&\cdot&\cdot&\cdot\\[4pt] 
 \cdot&\cdot&-1&\cdot&\cdot&\cdot\\[4pt] 
 \cdot&\cdot&\cdot&\frac{1}{2}&\cdot&\cdot\\[4pt] 
 \cdot&\cdot&\cdot&\cdot&\frac{-1}{2}&\cdot\\[4pt] 
 \cdot&\cdot&\cdot&\cdot&\cdot&\cdot \end{array}\right) &
J^8_{ \scriptstyle{ \textrm{ rep.6 }} } = \left( \begin{array}[c]{cccccc}
 \frac{1}{\sqrt{3}}&\cdot&\cdot&\cdot&\cdot&\cdot\\[4pt] 
 \cdot&\frac{1}{\sqrt{3}}&\cdot&\cdot&\cdot&\cdot\\[4pt] 
 \cdot&\cdot&\frac{1}{\sqrt{3}}&\cdot&\cdot&\cdot\\[4pt] 
 \cdot&\cdot&\cdot&\frac{-1}{2\sqrt{3}}&\cdot&\cdot\\[4pt] 
 \cdot&\cdot&\cdot&\cdot&\frac{-1}{2\sqrt{3}}&\cdot\\[4pt] 
 \cdot&\cdot&\cdot&\cdot&\cdot&\frac{-2}{\sqrt{3}}\end{array}\right)\\[25mm]
%
 I^+_{ \scriptstyle{ \textrm{ rep.6 }}} = \left( \begin{array}[c]{cccccc}
 \cdot&\sqrt{2}&\cdot&\cdot&\cdot&\cdot\\[4pt] 
 \cdot&\cdot&\sqrt{2}&\cdot&\cdot&\cdot\\[4pt] 
 \cdot&\cdot&\cdot&\cdot&\cdot&\cdot\\[4pt] 
 \cdot&\cdot&\cdot&\cdot&1&\cdot\\[4pt] 
 \cdot&\cdot&\cdot&\cdot&\cdot&\cdot\\[4pt] 
 \,\cdot\,&\,\cdot\,&\,\cdot\,&\,\cdot\,&\,\cdot\,&\,\cdot\, \end{array}\right)&
  V^+_{ \scriptstyle{ \textrm{ rep.6 }}} = \left( \begin{array}[c]{cccccc}
 \cdot&\cdot&\cdot&\sqrt{2}&\cdot&\cdot\\[4pt] 
 \cdot&\cdot&\cdot&\cdot&1&\cdot\\[4pt] 
 \cdot&\cdot&\cdot&\cdot&\cdot&\cdot\\[4pt] 
 \cdot&\cdot&\cdot&\cdot&\cdot&\sqrt{2}\\[4pt] 
 \cdot&\cdot&\cdot&\cdot&\cdot&\cdot\\[4pt] 
 \,\cdot\,&\,\cdot\,&\,\cdot\,&\,\cdot\,&\,\cdot\,&\,\cdot\, \end{array}\right)\\[25mm]
  U^+_{ \scriptstyle{ \textrm{ rep.6 }}} = \left( \begin{array}[c]{cccccc}
 \cdot&\cdot&\cdot&\cdot&\cdot&\cdot\\[4pt] 
 \cdot&\cdot&\cdot&1&\cdot&\cdot\\[4pt] 
 \cdot&\cdot&\cdot&\cdot&\sqrt{2}&\cdot\\[4pt] 
 \cdot&\cdot&\cdot&\cdot&\cdot&\cdot\\[4pt] 
 \cdot&\cdot&\cdot&\cdot&\cdot&\sqrt{2}\\[4pt] 
 \,\cdot\,&\,\cdot\,&\,\cdot\,&\,\cdot\,&\,\cdot\,&\,\cdot\,
\end{array}\right)
\end{longtable}


\section{Matrices for the fundamental representation {\bf 4} of SU(4)}
\label{app:SU4R4_matrices}
su(4) has three diagonal
generators $V^3$, $V^8$, and $V^{15}$.  The matrices for the
fundamental representation of SU(4) are given by:

\begin{longtable}{LLL}
  V^1 = \left( \begin{array}[c]{cccc}
   0&1&0&0\\ 1&0&0&0\\ 0&0&0&0\\ 0&0&0&0 \end{array}\right)&
  V^2 = \left( \begin{array}[c]{cccc}
   0&-i&0&0\\ i&0&0&0\\ 0&0&0&0\\ 0&0&0&0 \end{array}\right)&
  V^3 = \left( \begin{array}[c]{cccc}
   1&0&0&0\\ 0&-1&0&0\\ 0&0&0&0\\ 0&0&0&0 \end{array}\right) \\[12mm]
  V^4 = \left( \begin{array}[c]{cccc}
   0&0&1&0\\ 0&0&0&0\\ 1&0&0&0\\ 0&0&0&0 \end{array}\right)&
  V^5 = \left( \begin{array}[c]{cccc}
   0&0&-i&0\\ 0&0&0&0\\ i&0&0&0\\ 0&0&0&0 \end{array}\right)&
  V^6 = \left( \begin{array}[c]{cccc}
   0&0&0&0\\ 0&0&1&0\\ 0&1&0&0\\ 0&0&0&0 \end{array}\right)  \\[12mm]
  V^7 = \left( \begin{array}[c]{cccc}
   0&0&0&0\\ 0&0&-i&0\\ 0&i&0&0\\ 0&0&0&0 \end{array}\right)&
  \hspace{-2mm}V^8 = \frac{1}{\sqrt{3}}\left( \begin{array}[c]{cccc}
   1&0&0&0\\ 0&1&0&0\\ 0&0&-2&0\\ 0&0&0&0 \end{array}\right)&
  V^9 = \left( \begin{array}[c]{cccc}
   0&0&0&1\\ 0&0&0&0\\ 0&0&0&0\\ 1&0&0&0 \end{array}\right)  \\[12mm]
  V^{10} = \left( \begin{array}[c]{cccc}
   0&0&0&-i\\ 0&0&0&0\\ 0&0&0&0\\ i&0&0&0 \end{array}\right)&
  V^{11} = \left( \begin{array}[c]{cccc}
   0&0&0&0\\ 0&0&0&1\\ 0&0&0&0\\ 0&1&0&0 \end{array}\right)&
  V^{12} = \left( \begin{array}[c]{cccc}
   0&0&0&0\\ 0&0&0&-i\\ 0&0&0&0\\ 0&i&0&0 \end{array}\right)  \\[12mm]
  V^{13} = \left( \begin{array}[c]{cccc}
   0&0&0&0\\ 0&0&0&0\\ 0&0&0&1\\ 0&0&1&0 \end{array}\right)&
  V^{14} = \left( \begin{array}[c]{cccc}
   0&0&0&0\\ 0&0&0&0\\ 0&0&0&-i\\ 0&0&i&0 \end{array}\right)&
  \hspace{0mm}V^{15} = \frac{1}{\sqrt{6}}\left( \begin{array}[c]{cccc}
   1&0&0&0\\ 0&1&0&0\\ 0&0&1&0\\ 0&0&0&-3 \end{array}\right)\\[10mm]
\end{longtable}

\end{document}